\documentclass{aa}
\usepackage{epsfig}
\def\deg{^{\circ}}                                                  
\def\etal{{\rm et al.}\thinspace}

\def\bi{\bibitem{}}

\def\beb{\begin{thebibliography}{999}}
\def\eeb{\end{thebibliography}}
\def\bei{\begin{itemize}}
\def\eei{\end{itemize}}
\def\bef{\begin{figure}}
\def\eef{\end{figure}}
\def\ben{\begin{enumerate}}
\def\een{\end{enumerate}}
\def\beq{\begin{equation}}
\def\eeq{\end{equation}}
\def\ber{\begin{eqnarray}}
\def\eer{\end{eqnarray}}

\begin{document}
\title{Pulsar ``Drifting''-Subpulse Polarization:  No Evidence for Systematic 
Polarization-Angle Rotations}

\author{R. Ramachandran\inst{1} \and Joanna M. Rankin\inst{2}\thanks{On leave
from  Physics Dept., University of Vermont, Burlington, VT 05405 USA, email:
rankin@physics.uvm.edu} \and
B.W. Stappers\inst{1} \and M.L.A.Kouwenhoven\inst{3} \and A.G.J.van
Leeuwen\inst{3}}
\offprints{ramach@astro.uva.nl}
\institute{Stichting ASTRON, 7990 AA Dwingeloo, The Netherlands 
	\email{ramach@astro.uva.nl} \and 
	Sterrenkundig Instituut `Anton Pannekoek', 1098 SJ Amsterdam, 
	NL \email{jrankin@astro.uva.nl} \and
	Sterrenkundig Instituut, 3508 TA Utrecht, The Netherlands
	\email{A.G.J.vanLeeuwen@astro.uu.nl}}

\abstract{Polarization-angle density displays are given for pulsars
B0809+74 and B2303+30, which exhibit no evidence of the systematic
polarization-angle rotation within individual subpulses previously
reported for these two stars.  The ``drifting'' subpulses of both 
pulsars exhibit strikingly linear and circular polarization which 
appears to reflect the characteristics of two nearly orthogonally 
polarized emission ``modes''---along which the severe average-profile 
depolarization that is characteristic of their admixture at comparable 
overall intensities.  
}

\date{Received / Accepted }
\authorrunning{R. Ramachandran {\rm et al.}}
\titlerunning{Pulsar subpulse polarisation}
\maketitle

\section{Introduction}
\label{intro}
In this letter we provide straightforward, definitive evidence to 
the effect that pulsar ``drifting'' subpulses exhibit polarization 
reflecting virtually only the projected magnetic field direction.  
We find that the linear position angles (hereafter PAs) of such 
subpulses are oriented---like almost all other pulsar 
radiation---either parallel to or perpendicular to this direction.  
That is, drifting-subpulse polarization closely follows the 
rotating-vector model (hereafter RVM) first articulated by Radhakrishnan 
\& Cooke (1969) and by Komesaroff (1970).

Several influential papers, however, have suggested or reported precisely 
the opposite situation---that a characteristic rotation of the PA can 
be observed within some subpulses.  In perhaps the first report of 
individual pulse polarization---the oscilloscope images reproduced by 
Clark \& Smith (1969)---such an effect is clearly suggested, and other 
investigators writing within the first several years after the pulsar 
discovery (e.g., Lyne \etal 1971) emphasize the great variability of 
individual-pulse polarization in contrast to the usual great stability 
of the average or profile polarization, including the PA traverse.  
This latter compendium, along with Manchester's (1971) work, 
nonetheless demonstrated that most pulsars could at least partially 
be reconciled with the RVM, so that questions could be entertained 
regarding the manner in which the apparently disorderly subpulse 
polarization diverged from the more orderly average characteristics.
 
In this context, pulsar B0809+74 with its bright, regular, beautiful 
sequences of drifting subpulses has developed as the canonical example 
of extraordinary polarization behavior.  Both Lyne \etal and Manchester 
determined that the pulsar's average linear polarization at meter 
wavelengths was low, with a PA characterized by two shallow, negatively 
rotating segments offset by about 90$\deg$.  It was then interesting 
to assess whether its individual pulses exhibited rotations over and 
above what might be expected from changes in the projected magnetic 
field direction, or namely, whether their behavior seemed compatible 
with the RVM model above.

It was difficult, anywhere, in 1971 to measure the polarization of individual
pulses reliably, and in their now classic paper Taylor \etal used the NRAO
92-m telescope with a four-channel, 235-MHz polarimeter to record Stokes
parameters I, Q and U of sequences from pulsar B0809+74.  They noted that the
pulsar's polarization properties ``are closely linked with [its] bands of
drifting subpulses'' and that ``successive bands of subpulses displayed nearly
identical polarization behavior''.  Obvious also was that the individual
subpulses were much more highly polarized than the pulsar's average profile.
Such ``drift'' bands in this pulsar are, of course, regularly spaced by very
nearly 11 rotation periods, so Taylor \etal were able to average some 10
complete drifting cycles to improve their signal-to-noise (hereafter S/N) ratio.
This original and inventive technique then permitted the authors to remark that
``the position angle is moving with the subpulses and is not fixed on the
rotating star.''  The paper's fig.  1 indicates that most subpulses exhibit a
characteristic PA rotation irrespective of where they fall in longitude---that
is, most subpulses have a similar leading-edge PA which then rotates in a
consistent manner throughout its duration.  Taylor \etal's interpretation then
appears virtually inescapable.

Unfortunately, this conclusion is incorrect as we will show below, but the
authors of the above paper had no reasonable means of deducing this circumstance
at the time.  With 30-years hindsight we can now identify major clues to this
three-decade-old deception: a) the total rotation within each subpulse (in their
fig. 1) is usually close to 90$\deg$; and b) their resolution (between 3 and
$4\deg$) represents a signicant fraction of the subpulse width.  Indeed, only
slowly did evidence emerge to the effect that pulsar PAs tend to assume two
mutually orthogonal orientations.  Manchester \etal (1975) were the first to
give histograms suggesting a bimodal PA distribution in certain stars, but it
was not for a further five years that the generality of this phenomenon began to
be appreciated fully (Backer \& Rankin 1980).  By this time, however, most
interest in PSR B0809+74 and its drifting-subpulse polarization had
subsided.\footnote{Sadly, the one other report of systematic PA rotation during
subpulses in pulsar PSR B2303+30 (Gil 1992) is also incorrect; we now know that
the Arecibo Observatory polarimetry of the early 1970s was flawed by the use of
Gaussian-shaped IF filters.  This had the effect of correlating polarized power
at adjacent longitudes in an insidious manner.}  Only now are we beginning to
apprehend fully the crucial role played by the orthogonal polarization modes
within drifting-subpulse sequences.

\section{Polarization-Modal Structure of Conal Single Profiles}
\label{modalstruct}
We can now see clearly that conal single (${\bf S}_{\rm d}$) pulsar profiles are
strongly affected by the interaction of the two orthogonal polarization modes.
Of those stars known to have discernible {\it drifting} subpulses (e.g., see
Rankin 1986, 1993) every single one exhibits low fractional linear polarization
in the metre-wavelength region, and many show ``90$\deg$ flips'' (accompanied by
complete depolarization) at longitudes where the dominant mode switches---all of
which results in segmented or disorderly PA behavior bearing almost no
resemblance to any underlying geometrical (RVM) origin.

\begin{figure*}
\begin{center}
\epsfig{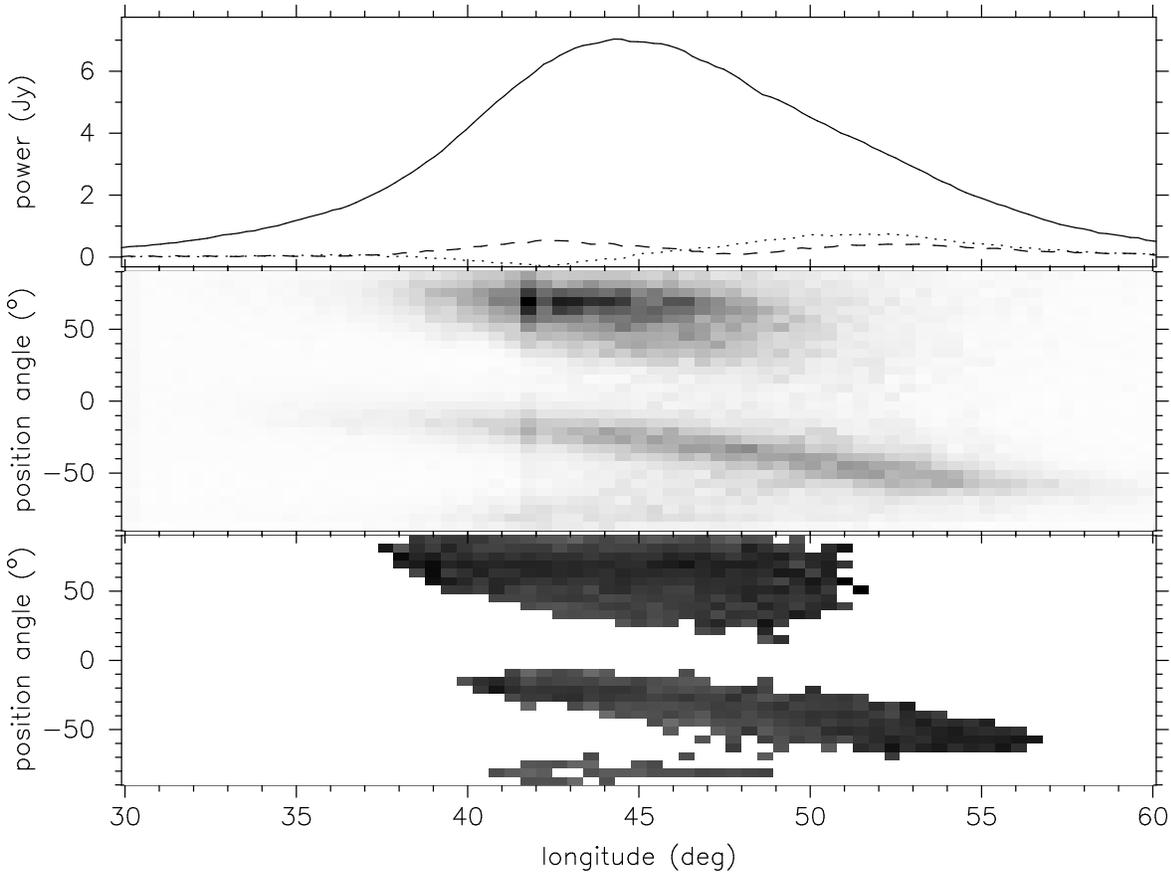}
\caption[]{Polarisation-angle-density diagram for PSR B0809+74 at 328 MHz. The
top panel gives the usual average Stokes parameters: total power $I$ (solid
line); average linear power $L$ (=$\sqrt{Q^2+U^2}$), corrected for the
statistical bias (dashed curve); and average circularly polarized power $V$
(dotted curve).  The usual box showing the resolution and off-pulse noise rms
level has not been plotted because it would be almost invisible.  The central
panel displays (in grey scale) the polarization position-angle distibution,
where the values in each pixel are weighted properly by the square of their S/N
level (see text).  Note the two distinct ``tracks'' corresponding to the two
orthogonal polarization modes (separated by about 90$\deg$). A sequence of
length 3600 pulses was used. The bottom panel shows (again in gray scale) the
fractional linear polarization distribution for those samples falling above a
five standard-deviation noise threshold and weighted as in the central
panel. The peak of the distribution is some 60\%.}
\label{fig:0809-328}
\end{center}
\end{figure*}

We can see this general behavior in all of these pulsars which 
have received adequate investigation: B0031$-$07 exhibits low 
linear polarization and PA ``flips'' below 1 GHz [Gould \& Lyne 
(1998); hereafter GL]; in PSR B0809+74 we see a similar behavior 
(Manchester 1971; Lyne \etal\ 1971); 0820+02 seems to show two 
``flips'' at 408 MHz (GL); 1923+04 has a single ``flip'' (Hankins 
\etal\ 2001); and in 2016+28, perhaps the best studied of all the 
${\bf S}_{\rm d}$ pulsars, we see consistent modal dominance 
effects near the center of its profile (e.g., see GL).  For 
2016+28 the polarization histograms in Backer \& Rankin (1980) 
and Stinebring \etal\ (1984)---see their figs. 18 and 31, 
respectively---leave no doubt about the characteristics of this 
overall modal behavior.  Studies of pulsar B0943+10 give a detailed, 
completely consistent example of the situation wherein one mode 
is consistently stronger than the other (Suleymanova \etal\ 1998; 
Deshpande \& Rankin 2001).  We also see this situation clearly 
documented in PSR B2303+30 (Backer \& Rankin 1980), and GL and 
Hankins \etal's work on 1540+06, 1612+07, 1923+04, 1940$-$12, 2110+27, 
and 2303+30 reveals compatible behavior.  Indeed, that 1923+04 
exhibits both behaviors (modal ``flips'' in some observations 
and not in others) suggests that modal power variations are 
probably common in ${\bf S}_{\rm d}$ stars, where they are often 
comparable in intensity---and slight changes in their relative 
amplitude would produce different average polarization 
effects.

\section{Observations and analysis}
Our 328-MHz observation of PSR B0809+74 was conducted with the 
Westerbork Synthesis Radio Telescope (WSRT), using its pulsar backend, 
{\tt PuMa}. WSRT is an east-west array with fourteen equatorially 
mounted 25-m dishes. For this observation (which was made on the 
26th November 2000), the delays between the dishes were compensated, 
and the signals were added `in phase' to construct an equivalent 
94-m single dish having a sensitivity of about 1.2 K Jy$^{-1}$. We
also calibrated the telescope array for polarisation measurements 
following the procedure given by Weiler (1973) and Weiler \& 
Raymond (1976; 1977). With a bandwidth of 10 MHz, centred around 
328 MHz, the signals were Nyquist-sampled, and Fourier transformed 
to synthesize a filterbank of 64 complex frequency channels. Stokes 
parameters were computed on-line in this frequency domain. Finally, 
after some averaging, 2-bit (4-level) data samples were recorded 
for all Stokes parameters and frequency channels, with a time
resolution of 819.2 $\mu$sec. During the offline analysis, we removed 
the interstellar-despersion delay between the signals of various 
frequency channels, producing multi-bit, floating-point sample values. 
For this purpose, we used a dispersion measure (DM) of 5.7513$\pm$0.0002 
pc cm$^{-3}$ as determined by Popov \etal (1987).

\begin{figure}
\begin{center}
\epsfig{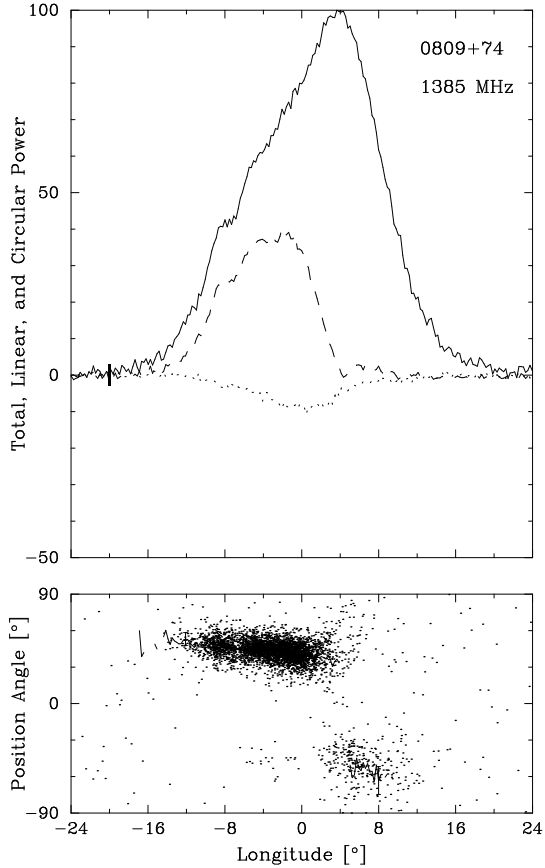}
\caption[]{Polarisation-angle histogram for PSR B0809+74 at
1365 MHz.  The top panel gives the usual average Stokes
parameters: total power $I$ (solid line); linear power
$L$ (=$\sqrt{Q^2+U^2}$) (dashed curve); and circularly
polarized power $V$ (dotted curve).  A box showing the
resolution and off-pulse noise rms level is just visible
at the left of the diagram.  The lower panel represents
the histogram of sample position angles with estimated
errors less than 30$\deg$ (see text),  together with the
superposed average PA traverse. Note the two distict
``tracks'' corresponding to the two orthogonal
polarization modes (separated by about 90$\deg$). A
2708-pulse sequence was used.}
\label{fig:0809-1420}
\end{center}
\end{figure}

\begin{figure}
\begin{center}
\epsfig{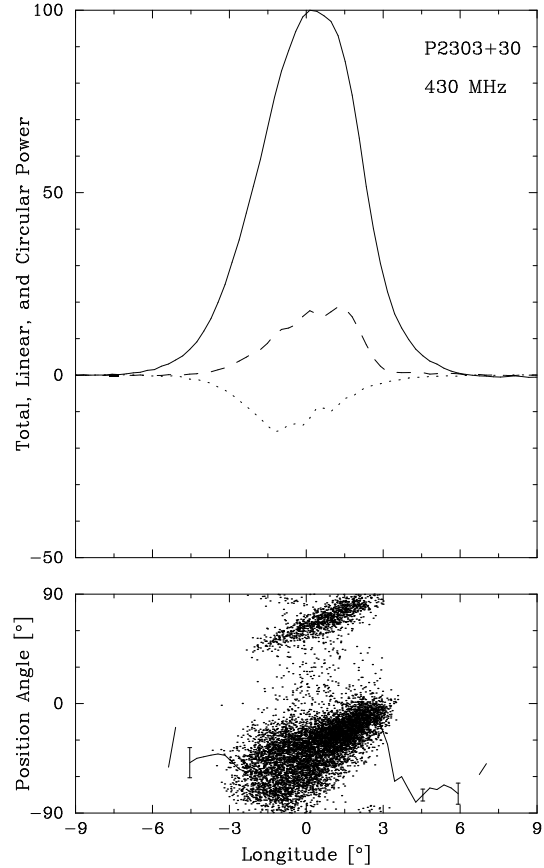}
\caption[]{PA histogram for PSR B2303+30 as in Fig. 2.
Again, note the two well separated modal PA ``tracks''.
The display used 2370 pulses and a polarization-angle
threshold of 8$\deg$.}
\label{fig:2303-430}
\end{center}
\end{figure}

The 1380-MHz observations were also conducted with the WSRT with its pulsar
backend, {\tt PuMa}. All the Stokes parameters were recorded as mentioned above,
with a bandwidth of 80 MHz and 512 frequency channels.

The 430-MHz observation of PSR B2303+30 was recorded at Arecibo 
on 15 October 1992, using the 40-MHz correlator, which ``dumped'' 
the ACF/CCF's of the right- and left-handed channel voltages at 
1206 $\mu$sec intervals.  Using 10-MHz bandwidth and retaining 32 
lags, dispersion delays were reduced to negligible values.  The 
time resolution was then about 0.$\deg$276 longitude (one rotation 
period equals 360$^{\circ}$ longitude).  After 3-level correction, 
the ACF/CCFs were calibrated and Fourier transformed to produce
Stokes parameters, which were then corrected for dispersion delays, 
Faraday rotation, and feed effects (see Suleymanova \etal 1998).

Figures~\ref{fig:0809-328} \& \ref{fig:0809-1420} (at 328 and 
1385 MHz, respectively) represent the first available polarization 
PA ``histograms'' for pulsar B0809+74.  The top panels in 
these two figures show the usual average polarization properties: 
total power (Stokes parameter $I$; solid curve), linear power, 
corrected for the statistical bias (dashed curve), and circular 
power ($V$; dotted curve). The middle panel of Fig.~\ref{fig:0809-328} 
and the bottom panel of Fig.~\ref{fig:0809-1420} then give the PA
distribution information: in the former as a grey scale with the 
values contributing to each pixel carefully weighted in order to 
maximize their significance; and in the latter as a display of 
sample values comprising a ``density plot''.  In the first figure, 
the values contributing to a pixel were weighted according the 
square of their S/N values---after taking care to estimate the PA 
error according to the procedure given in Rankin \& Rathnasree 
(1997; see footnote 11); whereas in the latter the PA values are 
given as ``dots'' corresponding to all data samples for which the 
error in the PA, $\sigma_\chi$ ($= \sigma_{\rm on}/L$) was 
less than 30$\deg$.  The on-pulse noise level $\sigma_{\rm on}$ 
was estimated as $\sigma_{\rm off}(I_{\rm sys}+L)/I_{\rm sys}$, 
where $I_{\rm sys}$ is the total power corresponding to the system 
temperature and $\sigma_{\rm off}$ is the standard deviation 
of the noise in Stokes parameter $I$ well away from the pulse window.  
(Clearly, $\sigma_{\rm off}/I_{\rm sys} = (2\tau \Delta\nu)^{-{1 
\over 2}}$, where $\Delta\nu$ is the total bandwidth and $\tau$ the 
effective integration time.).  Finally, the bottom panel of 
Fig.~\ref{fig:0809-328} gives the density of the fractional linear 
polarization, also as a grey scale---where a fully black pixel 
represents a value of about 60\%.  Here, in addition to S/N weighting 
as above, a threshold of 5 $\sigma_{\rm on}$ was imposed to 
improve the quality of the display.  This detailed analysis and 
display was possible for the 328-MHz sequence, because the average 
S/N was relatively high, but was not practical for the other two 
sequences. A display similar to Fig.~\ref{fig:0809-1420} 
for pulsar B2303+30 can be seen in Figure~\ref{fig:2303-430}. 

For both these pulsars the two parallel tracks---corresponding to 
the two orthogonal modes, separated by about 90$\deg$---can be very 
clearly seen.  For PSR B0809+74 at the lower frequency, the two 
modes have comparable power throughout much of the pulse, resulting 
in severe depolarization; whereas at the higher frequency we see 
little secondary-mode (herafter SPM) emission during the first half 
of the pulse, so that most of the depolarization occurs after the
peak of the average profile.  We can also see here that the 
PA-traverse rate associated with this SPM emission is -2$\deg/\deg$ 
or steeper, a fact that cannot readily be determined from the usual 
average polarimetry (e.g., the otherwise very well measured profiles 
recently of GL) as well as one having considerable importance for 
future efforts to understand the configuration of this pulsar's 
rotating subbeam system.  These PA ``tracks'' can now be assessed 
and interpreted according to the RVM of Radhakrishnan \& Cooke 
(1969)---and the initial evidence here is that they are nearly, 
but not precisely, orthogonal.

\begin{figure*}
\begin{center}
\epsfig{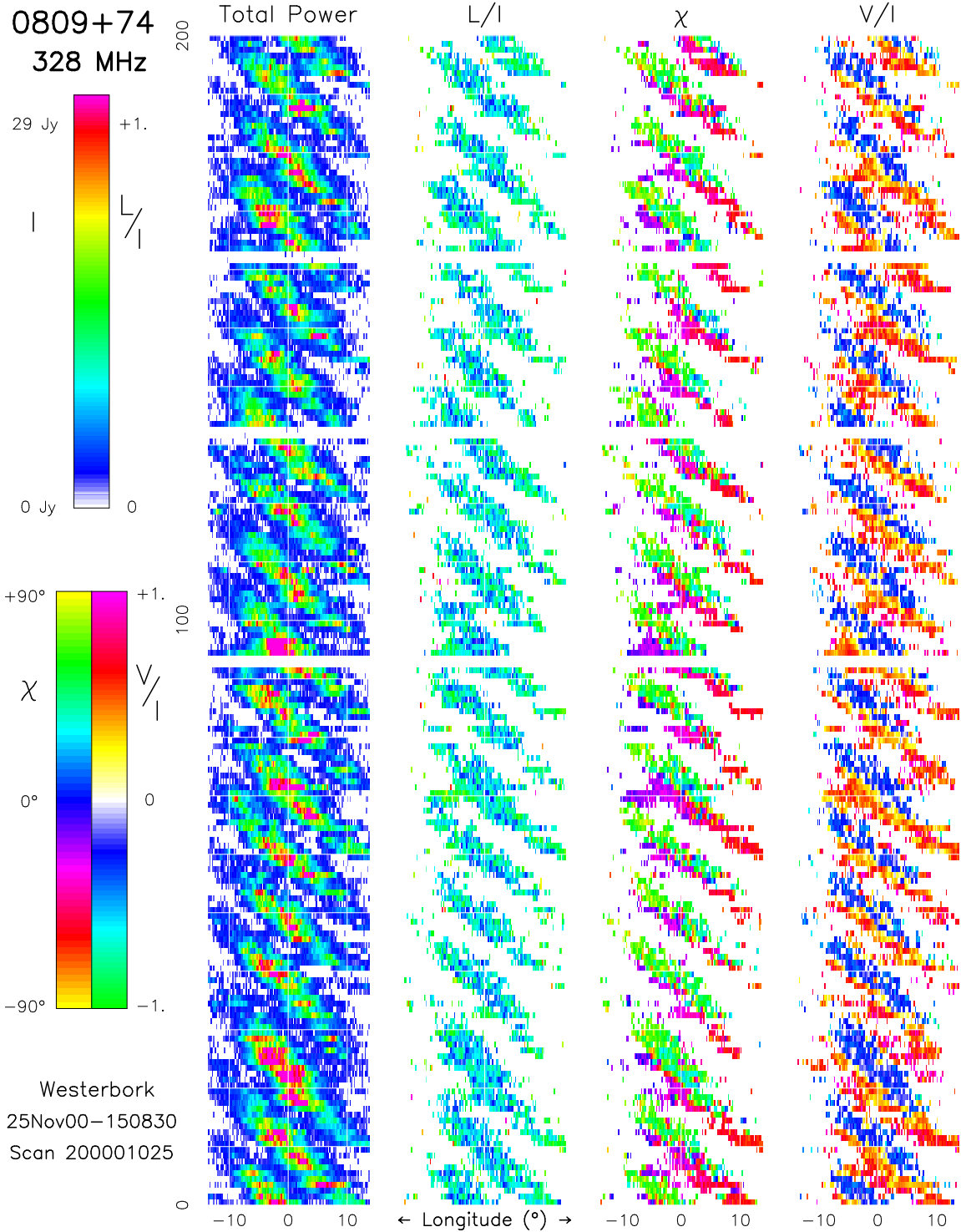}
\caption[]{Color display of a 200-pulse portion of 
the 328-MHz observation in Fig. 1.  The first column 
of the displays gives the total intensity (Stokes 
parameter $I$), with the vertical axis representing 
the pulse number and the horizontal axis pulse 
longitude, colour-coded according to the left-hand
scale of the top bar to the left of the displays.
The second and third columns give the corresponding
fractional linear polarisation ($L/I=\sqrt{Q^2+U^2}/I$)
and its angle ($\chi={1\over 2}\tan^{-1}{U/Q}$),
according to the scales at the top-right and
bottom-left of the left-most panel.  The last
column gives the fractional circular polarisation
($V/I$), according to the scale at the bottom-right
of the left-hand panel.  Plotted values have met 
a threshold corresponding to 2.5 standard deviations 
of the off-pulse noise level.  Note the strikingly 
modal character of the polarized power, with the 
linear assuming nearly orthogonal angles---color-coded 
as either chartreuse or magneta---and with the 
corresponding circular falling at about 40\% negative 
(purple) and positive (orange), respectively.  It is 
also notable that the PAs exhibit (apart from noise 
variations) virtually only these angle values, so that 
with a single pulse or subpulse, we see not primarily 
rotation,  but modal 90$\deg$ ``flips''. Can it be any 
wonder that polarization behavior of such remarkable 
complexity caused so much early confusion?}
\label{fig:0809-color}
\end{center}
\end{figure*}

We further note that while the PA histograms clearly exhibit the modal 
nature of the polarized power in pulsars 0809+74 and 2303+30, these
distributions remain remarkably complex.  The widths of the primary
polarization-mode (hereafter PPM) and SPM distributions is very
different.  When the two modal contributions have about equal power,
some of the radiation will be randomly polarised (as can be clearly
seen in the figures); whereas, when one dominates the other, its PA
determines the ensemble PA. Statistical theory discussions of this
phenomenon are rather complex (see Davenport \& Root 1958; Moran 1976;
and especially McKinnon \& Stinebring 1998).

Finally, the color polarization display in Figure~\ref{fig:0809-color} 
provides a completely different way of looking at the polarization 
characteristics of the subpulses.  The four columns give the total 
power, fractional linear polarization (corrected for statistical
bias), polarization angle, and fractional circular polarization of the
first 200 pulses of the 328-MHz sequence of Fig.~\ref{fig:0809-328}---all 
color-coded according to the respective color scales at the far 
left of the display.  Immediately, we see the drift bands---with 
the individual subpulses drifting from left to right (the pulses 
are numbered from the bottom to top), the three intervals of 
``null'' pulses, and the fairly consistent, about 40\% level of 
fractional linear polarization. More arresting, however, are the 
associated polarization angle and circular-polarization behavior, 
where we find two strongly preferred values for both the 
PAs---nearly orthogonal values coded chartreuse and magenta, 
respectively, and respective 40\% negative and positive fractional 
circular values, coded purple and orange.  

Careful inspection of the pulse-sequence PA behavior in the figure 
shows that there are very frequent 90$\deg$``flips''---usually 
from about -20$\deg$ (magenta) to +70$\deg$ (chartreuse)---but 
no good example of systematic rotation.  Of course, we see 
variations which may be partly modal and partly the effect of 
the noise, but were the pulsar subpulses are strongest (e.g., 
between number 18 and 27) the modal ``flips'' are clearly seen 
in every pulse.  We do see the longitude of these ``flips'' 
occuring at progressively earlier phases, so as to remain 
approximately parallel to the drift-band and thus within the 
subpulses as they ``drift''---and it was a smeared out signature 
of this phenomenon which Taylor \etal\ recorded 30 years ago.  

The full story is, however, much more complex: there is a 
progressive mixing of the two polarization modes as the subpulses 
drift across the profile ``window''---as subpulses at the extreme 
edges tend to exhibit only one of the modes.  In this context, 
the constancy of the fractional linear polarization is striking; 
indeed, one gets the impression that most of individual samples 
are not modally depolarized---only the contrasting (dark blue) 
ones which usually lie close to the PA ``flips''.  Also intriguing
is the circular, which is highly correlated with the modal linear 
power, while exhibiting a slightly displaced distribution along 
the drift bands.  We will return to these questions in much more 
detail in subsequent papers on 0809+74.

\section{Discussion}

Polarisation position-angle displays for PSRs B0809+74
and PSR B2303+30, computed from well measured sequences 
and given above, show clearly that the PAs trace out two 
well defined, nearly parallel trajectories (separated by 
approximately 90$\deg$) as a function of longitude.  As 
such, they individually trace curves which are completely 
compatible (within their errors) with the Radhakrishnan 
\& Cooke (1969) RVM, and thus entirely exclude the large 
($\sim90\deg$), extraordinary, subpulse-related PA 
rotations previously reported for these two stars. If 
such systematic rotations occurred, the character of the 
distributions would have to be very different---and the 
Taylor \etal\ result---about 90$\deg$ PA rotation across 
subpulses occurring over a subpulse-width-order interval 
of longitude {\it is} completely explained by the the 
prominent, stongly longitude-segregated regions of modal 
power, together with their subpulse-width-order instrumental 
resolution.  Of course, we can say nothing about any small 
extraordinary PA rotation within the respective PPM and 
SPM ``tracks''---but this question is well beyond the 
limited scope of our paper.   

Nonetheless, we also see here that the character of the 
modal emission is exceedingly complex, entailing at least 
four factors: a) its overall angular and/or temporal 
distribution---probably following from its generation 
within a rotating subbeam system, b) our viewing geometry, 
which determines the manner in which the profile ``window'' 
weights the various contributions, and c) the level of 
instrumental noise. The PA distributions for PSR B0809+74's 
two modes at 328 MHz, for instance, are quite different; 
the PPM's PA width is much broader than the SPM and their 
separation may not be just 90$\deg$.  For all these reasons 
the depolarization in this pulsar is very complex, such 
that the individual pulse polarization---which is typically 
40\% linear {\it and} circular---is reduced in their 
aggregation to hardly 5 and 10\%, respectively.  
  
Consequently, it should perhaps neither be surprising that 
PSR B0809+74's subpulses appeared to exhibit the extraordinary 
``PA rotation'' on the basis of early observational and 
analytical methods, nor that it has taken fully 30 years 
to discern that this interpretation was incorrect. Thus, 
we now hope and trust that this truer, more universal 
characterization of the pulsar's polarized emission will 
facilitate broader and more comprehensive physical explanations 
of its overall nature.

\acknowledgements
We gratefully acknowledge the contributions of our 
referee, Jean Eilek, whose comments and criticisms 
greatly assisted us in clarifying our presentation 
of the polarimetric material above.  One of us (JMR) 
also wishes to acknowledge the support of US National 
Science Foundation Grant AST 99-87654.  The Westerbork 
Synthesis Radio Telescope is administered by ASTRON 
with support from the Netherlands Foundation for 
Research in Astronomy.  Arecibo Observatory is operated 
by Cornell University under contract to the US NSF.  

\beb
\bi Backer, D. C., Rankin, J. M. 1980, ApJS, 42, 143.
\bi Clark, R. R., Smith, F. G. 1969, Nature 221, 724.
\bi Davenport, W. B., Root, W. L. 1958, ``Random signals and noise'' (New
    York: McGraw Hill)
\bi Deshpande, A. A., Rankin, J. M. 2001, MNRAS 322, 488.
\bi Gil, J. A. 1992, A\&A 256, 495.
\bi Gould, D. M., Lyne, A. G. 1998, MNRAS 301, 253.
\bi Hankins, T. H., Rankin, J. M. \& Eilek, J. A. 2001, preprint.
\bi Komesaroff, M. M. 1970, Nature, 225, 612
\bi Lyne, A. G., Smith, F., Graham, D. 1971, MNRAS 153, 337.
\bi Manchester, R. N. 1971, ApJS, 23, 283.
\bi Manchester, R. N., Taylor, J. H., Huguenin, G. R. 1975, ApJ 196, 83. 
\bi McKinnon, M. M., Stinebring, D. R. 1998, ApJ, 502, 883
\bi Moran, J. M. 1976, Meth. Exp. Physics, 12C, 228
\bi Popov, M. V., Smirnova, T. V., Soglasnov, V. A. 1987, Sov. Astr., 31, 529
\bi Radhakrishnan, V., Cooke, D. J. 1969, ApLett, 3, 225
\bi Rankin, J. M. 1986, ApJ, 301, 901.  
\bi Rankin, J. M. 1993, ApJS, 85, 145
\bi Rankin, J. M. \& Rathnasree, N. 1997, JAA, 18, 91
\bi Stinebring, D. R., Cordes, J. M., Rankin, J. M., Weisberg, J. M., 
    Boriakoff, V. 1984, ApJS, 55, 247.
\bi Suleymanova, S. A., Izvekova, V. A., Rankin, J. M., Rathnasree, N. 
    1998, JAA 19, 1.
\bi Taylor, J. H., Huguenin, G. R., Hirsch, R. M., Manchester, R. M. 
     1971, ApLett 9, 205.  
\bi Weiler, K. W. 1973, A\&A, 26, 403
\bi Weiler, K. W., Raimond, E. 1976, A\&A, 52, 397
\bi Weiler, K. W., Raimond, E. 1977, A\&A, 54, 965
\eeb
\end{document}